# On-off intermittency of thalamo-cortical oscillations in the electroencephalogram of rats with genetic predisposition to absence epilepsy

Evgenia Sitnikova[1,]*, Alexander E. Hramov[2], Alexey Ovchinnikov, Alexey A. Koronovskii[3]

[1] Institute of the Higher Nervous Activity and Neurophysiology of Russian Academy of Sciences, Butlerova str., 5A, Moscow, 117485, Russia, e-mail: eu.sitnikova@gmail.com

[2] Faculty of Nonlinear Processes, Saratov State University, Astrakhanskaya str., 83, Saratov, 410012, Russia, e-mail: hramovae@gmail.com

[3] Faculty of Nonlinear Processes, Saratov State University, Astrakhanskaya str., 83, Saratov, 410012, Russia, e-mail: alexey.koronovskii@gmail.com

* - send correspondence to e-mail: eu.sitnikova@gmail.com Institute of the Higher Nervous Activity and Neurophysiology of Russian Academy of Sciences, Butlerova str., 5A, Moscow, 117485, Russia

ABSTRACT

Spike-wave discharges (SWD) are electroencephalographic hallmarks of absence epilepsy. SWD are known to originate from thalamo-cortical neuronal network that normally produce sleep spindle oscillations. Although both sleep spindles and SWD are considered as thalamo-cortical oscillations, functional relationship between them is still uncertain. The present study describes temporal dynamics of SWD and sleep spindles as determined in long-term EEG recordings in WAG/Rij rat model of absence epilepsy. It was found that non-linear dynamics of SWD fits well to the law of ‗on-off intermittency'. Typical sleep spindles that occur during slow-wave sleep (SWS) also demonstrated ‗on-off intermittency' behavior, in contrast to high-voltage spindles during intermediate sleep stage, whose dynamics was uncertain. This implies that both SWS sleep spindles and SWD are controlled by a system-level mechanism that is responsible for regulating circadian activity and/or sleep-wake transitions.

Abbreviations:
SWD - spike-wave discharges
SWS - slow-wave sleep
IS - intermediate sleep stage

1. INTRODUCTION

Sleep spindles are among the most numerous spontaneous oscillations that are abundantly present in electroencephalograms (EEG) during non-REM sleep in humans and in animals [reviewed in De Gennaro and Ferrara, 2003]. Sleep spindles can be recorded at the cortical surface, and also in the thalamus and in some other brain structures as brief episodes of 9-14 Hz oscillations (so-called spindle sequences). It comes from basic electrophysiological studies that sleep spindle oscillations are triggered by reticular thalamic nuclei and then spread throughout the thalamus and propagated to the cortex [for refs see: Destexhe and Sejnowski, 2001; Steriade, 2003]. Thalamo-cortical neuronal circuit, which normally produces sleep spindles, under certain conditions[1] could give rise to epileptic spike-wave discharges (SWD). SWD are electroencephalographic hallmarks of generalized idiopathic epilepsies, such as absence epilepsy and other syndromes. Typical 3 Hz SWD could be seen in childhood absence epilepsy (‗pyknolepsy') [Panayiotopoulos, 1997].

Reciprocal relationship between sleep spindles and SWD was first established in pharmacological models of epilepsy. It was found that spindle oscillations were gradually transformed into SWD after local or systemic applications of penicillin. In high doses, penicillin acted as GABAergic antagonist and caused an increase of cortical excitation [Gloor, 1968; Steriade et al., 1993; Kostopoulos, 2000]. This ‗cortico-reticular' theory is still well accepted; despite the fact that typical GABAergic antagonists, such as picrotoxin or bicuculline, are not capable of inducing a transformation of spindle-like oscillations into spike-wave discharges.

The biggest arguments against direct relationship between spindles and SWD have been obtained in drug-free animals with genetic predisposition to absence epilepsy [Pinault et al., 2006; Sitnikova, 2010]. These studies were performed in two rat strains (GAERS and WAG/Rij rats) that developed spontaneous SWD and have been widely used in a field of basic research as reliable models of human absence epilepsy [Marescaux et al., 1992; Coenen and van Luijtelaar, 2003]. For example, our investigations in WAG/Rij rats showed that spindle activity and SWD (i) displayed different time-frequency characteristics as measured in cortex and thalamus [Sitnikova et al., 2010], (ii) were underlain by different neuronal processes and different neurotransmission mechanisms [Sargsyan et al., 2007]. Therefore, sleep spindles and SWD have been considered as autonomous EEG phenomena and straightforward relationship between then is doubtful [Sitnikova and van Luijtelaar, 2009; Sitnikova et al., 2009; Sitnikova, 2010].

---

[1] This can happen if neuronal network synchronization is too high (hyper-synchronization), or in case if cortical neurons exhibit too strong excitatory responses to thalamic volleys (hyper-excitation).

On the other hand, it is well known that sleep spindles and spontaneous SWD (but not pharmacologically induced seizures) are characterized by similar temporal distribution across sleep-waking cycle. In particular, both EEG events are predominant in drowsy state and in transition from wakefulness to sleep [e.g., Drinkenburg et al, 1991; Steriade, 2003; van Luijtelaar and Bikbaev, 2007]. Sleep spindles are abundant during slow-wave sleep (SWS); and circadian dynamics of SWD also positively correlates with the dynamics of SWS, as it has been demonstrated in WAG/Rij rat model of absence epilepsy [Drinkenburg et al, 1991]. Indeed, absence epilepsy is a sleep-related disorder, and SWD appear more often appear when the level of vigilance is low (relaxation, drowsiness, sleep). Altogether this implies that occurrence of sleep spindles and SWD in EEG could be controlled by a common circadian timing mechanism that regulates sleep-wake cycle. In order to elucidate this point, we examined non-linear aspects of temporal dynamics of sleep spindles and SWD in long-term EEG records in freely moving WAG/Rij rats.

Our study was focused on thalamo-cortical oscillations, i.e., SWD and sleep spindles. Besides typical sleep spindles that occur during slow-wave sleep (SWS spindles) we distinguished high-voltage spindles during intermediate sleep stage (IS-spindles). The latter seemed to be an intermediate form between SWS spindles and SWD. In total, we examined three kinds of oscillatory patterns, SWD, SWS-spindles and IS-spindles, in respect to their intrinsic time-frequency structure and global dynamics. Time-frequency analysis was performed by means of the continuous wavelet analysis, and nonlinear dynamics was statistically evaluated according to the theory of intermittency (see «Methods» for details).

2. METHODS

*2.1. Animals and EEG recording*

EEGs were recorded in five male WAG/Rij rats (one year old, body weigh 320-360 g). Animals were born and raised at the laboratory of Biological Psychology, Donders Institute for Brain, Cognition and Behavior of Radboud University Nijmegen (The Netherlands). The experiments were conducted in accordance with the legislations and regulations for animal care and were approved by the Ethical Committee on Animal Experimentation of the Radboud University Nijmegen. Distress and suffering of animals was kept to a minimum.

A recording electrode was implanted epidurally over the frontal cortex for the reason that SWD and spindles showed their amplitude maximum in this zone (coordinates: AP +2 mm and L 2.5 mm relatively to the bregma). Ground and reference electrodes were placed over the two symmetrical sides of the cerebellum. EEG recordings were made in freely moving rats during dark period of the light-dark cycle continuously for 5-7 hours. EEG signals were fed into a multi-channel differential amplifier via a swivel contact, band-pass filtered between 0.5-100 Hz, digitized with 1024 samples/second/per channel (CODAS software).

*2.2. Description of EEG patterns*

Sleep spindles were selected visually in accordance to the guidelines for clinical electroencephalography [Rechtschaffen and Kales, 1968; IFS, 1974] by taking into account specificity of sleep spindle activity in rats [Terrier and Gottesmann, 1978; Steriade et al. 1993; van Luijtelaar, 1997]. Sleep spindles in EEG were recognized as a sequence of 8-14 Hz waves with minimal duration of 0.3 sec (e.g., at least three wave cycles in a sequence). Sleep spindles had waxing-waning morphology and were characterized by twofold increase in amplitude as compared to EEG background. In order to facilitate expert's estimations, EEG records were additionally band-pass filtered 5-15 Hz. Visual spindle detections were double checked and verified by another expert.

SWD appeared in the EEG as a sequence of repetitive high-voltage negative spikes and negative waves that lasted longer than 1 sec; amplitude of SWD exceeded background more than three times [van Luijtelaar and Coenen, 1986]. SWD were first detected automatically with EEG slope method (Dr. PLC van den Broek) which provided about 90% of correct detections and then verified by two experts.

## 2.3. EEG analysis: continuous wavelet transform

Continuous wavelet transform (CWT) was used for high-resolution representation of EEG signal in time and frequency domains [Torresani, 1995; Koronovskii and Hramov, 2003]. Besides that, the CWT was used here in order to differentiate between bursting activity ('on' state) and the rest EEG ('off' state, in terms of 'on-off intermittency', see below).

The CWT, $W(s,\tau)$, was obtained by convolving the EEG signal, $x(t)$, with wavelet basis function, $\phi_{s,\tau}$ (Eq. 1,2)

$$W(s,\tau) = \int_{-\infty}^{+\infty} x(t) \phi^*_{s,\tau}(t) dt \quad (1), \quad \text{'*' denotes complex conjugation}$$

$$\phi_{s,\tau}(t) = \frac{1}{\sqrt{s}} \phi_0\left(\frac{t-\tau}{s}\right) \quad (2), \quad \text{'s' is time scale, and } \tau \text{ is time shift of wavelet transform}$$

The time scale (or simply 'scale') is inversely proportional to frequency.

Complex Morlet wavelet (Eq.3) was used as basis function, $\phi_{s,\tau}$ in Eq. 2.

$$\phi_0(\eta) = \frac{1}{\sqrt[4]{\pi}} e^{j\omega_0 \eta} e^{\frac{-\eta^2}{2}}. \quad (3), \quad \omega_0 \text{ is an empirically defined parameter}$$

In the present study, we used $\omega_0 = 2\pi$ that resulted in optimal time-frequency resolution of the transformed EEG signal. Time scales $s$ were converted into Fourier frequencies $f_s$ using the simple formula: $f_s = 1/s$ (this formula is true for $\omega_0 = 2\pi$, and it could be more sophisticated for other $\omega_0$ values).

The modulus of the CWT, $|W(s,t_0)|$, characterizes intensity of time scales $s$ in the transformed signal, $x(t)$, at the given time moment $t_0$.

Wavelet energy distribution $E(f_S)$ in Eq. 4 represents a wavelet spectrum, or scalogram, similar to spectrogram (Fourier power spectrum).

$$E(f_s) = \frac{1}{T} \int_0^T |W(t, f_S)|^2 \, dt. \quad (4)$$

Distribution of wavelet energy over the frequency scale was visually inspected in order to assess frequency characteristics of sleep spindles and SWD.

## 2.4. Analysis of non-linear dynamics

For the analysis of temporal dynamics, it was necessary to choose an appropriate non-linear model that take into account intrinsic features of sleep spindles and SWD. Below are the most relevant features.

Synchronization. It is known that both sleep spindles and SWD are manifestations of neuronal synchronization in EEG and they are correspond to neuronal bursts on cellular level [Steriade, 2003]; whereas the rest EEG is less synchronized.

Oscillatory activity. Spindle events and SWD are oscillatory EEG phenomena, but they are distinguished in frequency domain [e.g., Sitnikova et al., 2009]. This suggests that sleep spindles and SWD are underlain by different neuronal mechanisms and that their dynamics could be different.

Suddenness/predictability. It is well known that absence seizures in human patients are characterized by an abrupt the onset and termination [Panayiotopoulos, 1997], and SWD appears in EEG unpredictably.

In model systems, similar features are common for intermittent behavior [e.g., Perez Velazquez et al., 1999], more specifically, *on-off* intermittency [Nagai et al., 1996; Hramov et al., 2005].

It is known that *on-off* intermittency appears in a system that continually alternate between brief episodes of bursting activity (_on'-state) and quiescent _off'-state [Heagy et al., 1994; Kim C.-M., 1997].

In the present study we used an analytical approach that has been previously developed in order to describe non-linear dynamics of SWD in terms of *on-off* intermittency (for more details see [Koronovskii et al., 2006; Hramov et al., 2006]). Briefly, this method relied on measures of wavelet energy distribution in EEG. Elevations of wavelet energy in characteristic frequency bands were characteristic for episodes oscillatory activity in EEG (i.e., episodes of EEG synchronization, either SWD or spindle events), which were considered as the _on'-state of intermittent behavior. Periods between events, i.e., desynchronized EEG, represented the _off'-state. In the current study, both types of oscillatory EEG patterns, SWD and sleep spindles, were regarded as the _on'-state of thalamo-cortical oscillatory system, under the assumption that SWD and sleep spindles represented two different types of _on'-state.

*Off*-phases were marked in the full-range EEG, and their durations were measured. This analysis was performed separately for SWD and spindles as it is illustrated in Fig.1. $L$ - duration of *off*-phases of intermittency, i.e., periods between two consequent SWDs (or between spindle events). Then, we examined the dependence of the number of *off*-phases, $N(L)$, on their duration, L. Finally, we tested our data for power-law distribution, using the following function:

$N(L) = \beta L^{\alpha}$, (5) where $\alpha$ is an exponent, and $\beta$ is normalization factor

Power-law approximation was performed in logarithmic space, in which power-law representation was obtained in normal coordinates, using least-squares minimization criteria. The value of parameter $\alpha$ in Eq.5 was of particular interest, inasmuch as $\alpha = -3/2$ is a distinctive feature for a system with *on-off* intermittency [Heagy et al., 1994; Cabrera et al, 2002; Hramov et al., 2006].

The value of power _$\alpha$' was defined by minimizing the root-mean square error, $\varepsilon$, of empirically calculated distributions with different bin widths, $\Delta L$. Figure 2 (upper plot) shows the dependence between bin widths _$\Delta L$' and the exponent of power-law function, $\alpha$, characterizing distribution of SWS sleep spindles. It can be seen that the value of _$\alpha$' decreased from –0.6 to –2.3 with the increase of _$\Delta L$'. The root-mean square error was minimal for bin width ~4 (Fig.2, bottom plot), and this corresponded to the power law with the exponent $\alpha=-3/2$.

## 3. RESULTS
### 3.1. *Wavelet analysis of SWD and sleep spindles*

Time-frequency analysis of SWD and sleep spindles was performed with the aid of Morlet-based CWT. Wavelet spectrum of SWD usually displayed two dominant frequency components (Fig.3A): ~8-10 Hz and its harmonics in 16-20 Hz. Sharp spike elements in SWD can be recognized as high-frequency bursts in wavelet spectrum (above 20 Hz).

Time-frequency portrait of sleep spindles was heterogeneous. The most sharp differences were found between sleep spindles during slow-wave sleep (SWS spindles) and intermediate sleep state (IS spindles). Wavelet spectrum of SWS sleep spindles (Fig.3B) revealed a mixture of low frequency components (up to 10 Hz) and local elevations in frequencies between 10 and 15 Hz. Wavelet spectrum of IS spindles, in contrast to SWS spindles, showed a remarkable peak in ~10 Hz and more power in theta.

Figure 4 demonstrates the instantaneous wavelet energy distribution, $E(f_s)$ in Eq.3, as measured in IS and SWS sleep spindles. The following distinctions can be noted between SWS and IS spindle events. First, SWS spindle showed less wavelet energy as compared to IS spindle. Second, IS spindle displayed clear peaks in delta (2-4 Hz), theta (6-8 Hz) and alpha (10-14 Hz) frequency bands, in contrast to SWS spindle, whose wavelet power was centered around delta and alpha frequencies, but not in theta. IS spindles lasted longer than SWS spindles (0.8±0.2 s versus 2.5±0.9 s). Time-frequency properties of IS spindles were somewhat similar to SWD and to SWS spindles. In general, oscillatory pattern of IS spindles combines some features of SWD and sleep spindles.

## *3.2. Non-linear dynamics of SWD and sleep spindles*

Temporal dynamics of SWD and sleep spindles was identified based on statistical analysis of time periods between consequent SWD and between consequent sleep spindles ($L$ intervals, corresponding to the *off*-phase of intermittent behavior, see Methods for details). Distribution histograms of between-SWD and between-SWS-spindle intervals, $N(L)$, were plotted in log-log scale (Fig. 5). It was found that distributions of both between-SWD and between-spindle intervals were close to the straight line, corresponding to the power law with the exponent −3/2: $N(L)=L^{\alpha}$, where $\alpha = -3/2$. As it has been already mentioned this is a characteristic feature of *on-off* intermittency.

Although the number of SWD (and SWS spindles) varied from rat to rat, these EEG events showed −3/2 power law distributions in all experimental animals. For IS spindles, distributions of between-spindle intervals displayed a good approximation to the power law with the exponent −1. This was rather uncertain dynamics that obviously differed from the *on-off* intermittent behavior of SWS-spindles and SWD.

It is important that *on-off* intermittent behavior was found in both SWD and SWS spindles, despite the fact that these EEG events differed in respect to their time-frequency profiles (Fig. 3), and they are also underlain by differed mechanisms of neuronal network synchronization.

## 4. DISCUSSION

The present paper demonstrates that occurrence of spindle events and seizure activity in EEG is non-random, and that temporal dynamics of SWS-spindles and SWD can be characterized as *on-off* intermittency. This was found in all experimental animals and may imply a powerful intrinsic mechanism that governs the occurrence of epileptic activity and spindle events during slow-wave sleep. The proper context for understanding the nature of this mechanism can be provided the fact that SWS-spindles and SWD are produced by the same thalamo-cortical neuronal network mechanisms, and they appear in the same behavioral state, i.e., low vigilance state, such as drowsiness and sleep. This could also imply that both SWS sleep spindles and SWD are controlled by a higher system-level mechanism that regulates circadian activity and/or sleep-wake transitions (i.e., neuromodulatory systems, reticular activating system).

We have already reported that absence seizures in WAG/Rij rats demonstrated *on-off* intermittent behavior [Koronovskii et al., 2006; Hramov et al., 2006]. It is not surprising, because intermittency has already been described in spontaneously occurring synchronized EEG oscillations in humans. In particular, Gong et al. (2007) found that dynamics of synchronized episodes of alpha activity in human EEG had the property of type-I intermittency. The authors assumed that *"this kind of dynamics enables the brain to rapidly enter and exit different synchronized states, rendering synchronized states metastable"* (p. 011904-1). Based on our results, we can add that intermittent behavior may characterize rapid switch between oscillatory and non-oscillatory brain activity; this mechanism might help to avoid unstable oscillatory regimes that could emerge in transition from one stable state to another. Thalamo-cortical oscillations, i.e., spike-wave discharges and sleep spindles, reflect global neuronal network synchronization. *On-off* intermittency is known to appear at the boundaries of synchronization behavior [Kim C-M, 1997; Boccaletti et al, 2000; Hramov et al, 2005]. In other words, the process of appearance/disappearance of synchronization patterns (*on*-states) is negligible in systems with intermittent behavior. It seems likely that disruption of synchronous activity in thalamo-cortical neuronal network (desynchronization after SWD or sleep spindle events) may be accomplished through on-off intermittent mechanism.

Another conclusion that we draw from the current results is that SWS spindles differed from IS spindles in respect to their time-frequency properties and circadian dynamics. To some extent IS spindles are similar to SWD, but their temporal dynamics is totally different. Perhaps timing mechanism of IS spindles is distinct in nature from the putative vigilance controlling mechanism that regulates occurrence of SWS spindles and SWD. Alternatively, IS spindles might not involve thalamo-cortical neuronal network, which is known to produce SWS spindles and SWD.

Based on literature and our own data, we tend to consider IS spindles as pro-epileptic EEG phenomena for several reasons. First, it is known that IS spindles in non-epileptic rats appear less frequently than in epileptic animals; and also duration of IS spindles in non-epileptic rats was three times shorter as compared that in WAG/Rij rats [Gandolfo et al., 1990]. Second, theta frequency component was found in IS spindles, but it was absent in SWS spindles (current observation). Short episodes of theta-oscillations are known to immediately precede SWD in WAG/Rij rats and GAERS [Pinault et al., 2001]. Perhaps, IS spindles are somewhat abortive SWD, which appear in a period when thalamo-cortical network is not feasible to sustain SWD.

From the technical point of view, the current paper extends nonlinear approach to the description and explanation of temporal dynamics of spontaneous oscillatory events in EEG and provides the new wavelet-based algorithm for studying several oscillatory patterns in one EEG. This clearly non-random intermittent behavior of spontaneous oscillatory synchronous activity in EEG may imply a powerful intrinsic mechanism controlling occurrence of epileptic discharges and normal sleep activity. Further studies might be of clinical interest and may help to build up effective tools for predicting epileptic activity and for prognosis of seizure remission.

**ACKNOWLEDGEMENTS**

We highly appreciate Dr. Gilles van Luijtelaar for providing experimental facilities and WAG/Rij rats in Donders Institute for Brain, Cognition and Behavior. A.A.K. thanks ―Dynasty‖ Foundation for financial support. This study was supported by Russian Foundation for Basic Research (RFBR, 09-04-01302).

FIGURE CAPTIONS

Figure 1. An example of EEG recorded in WAG/Rij rat. On the upper plate, spike-wave discharges (SWD) can be seen as high-voltage bursts. On the bottom plate, sleep spindles are shown in filled squares with larger time resolution. ‚L' is a period between two consequent events (either SWD or sleep spindles).

Figure 2. Results of goodness-of-fit tests for power-law distribution with the exponent ‚α' (SWS sleep spindles). Root-mean square error, ε, was calculated between theoretical power law (Eq.5) and empirical distribution of between-spindle intervals (L) for different bin widths (ΔL).

Figure 3. Continuous wavelet transform of epileptic discharges (SWD) and sleep spindles in EEG as recorded in WAG/Rij rat. Both types of investigated EEG phenomena revealed complex structure in frequency domain with dominant frequency in alpha range (see text for details).

Figure 4. Distribution of wavelet energy $E(f_s)$ across frequencies as measured in SWS and IS sleep spindle events.

Figure 5. Distribution histograms of between-SWD and between-spindle intervals (L) in log-log scale. In slow-wave spindles (SWS-spindles) and SWD, distributions of L-intervals are best approximated to the power law with the exponent −3/2 in all individuals (ID=1-5). This regularity is an indication of on-off intermittency. In intermediate sleep spindles (IS-spindles), distribution of between-spindle intervals fits the power law with the exponent −1 (uncertain dynamics).

**Figure 1**

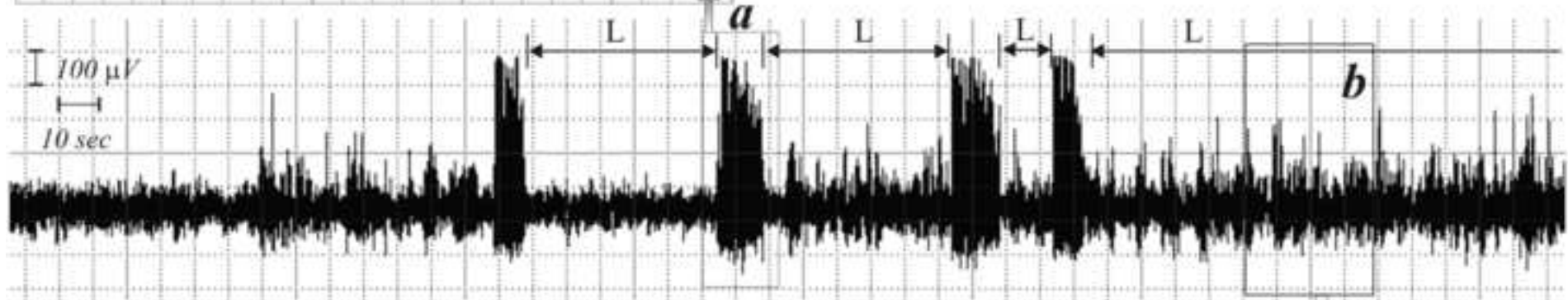
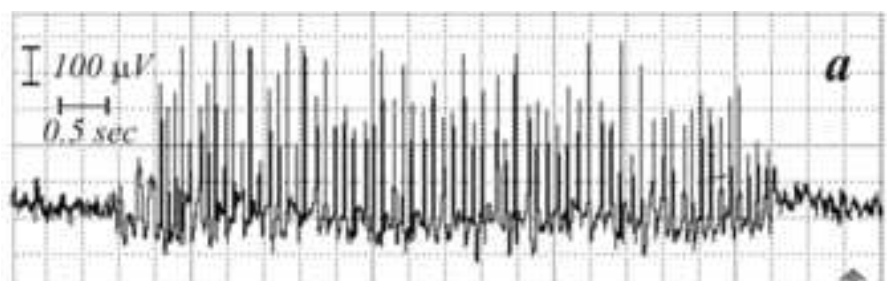
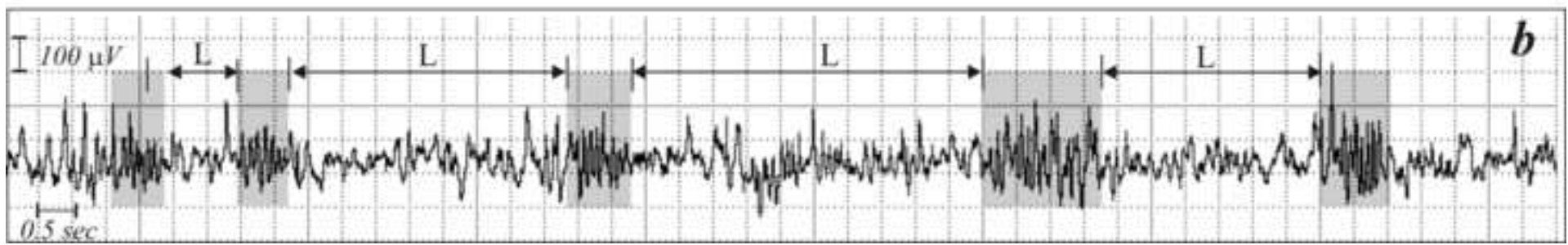

A. *Spike-wave discharges, SWD*

B. *Slow-wave sleep spindles*

**Figure 2**

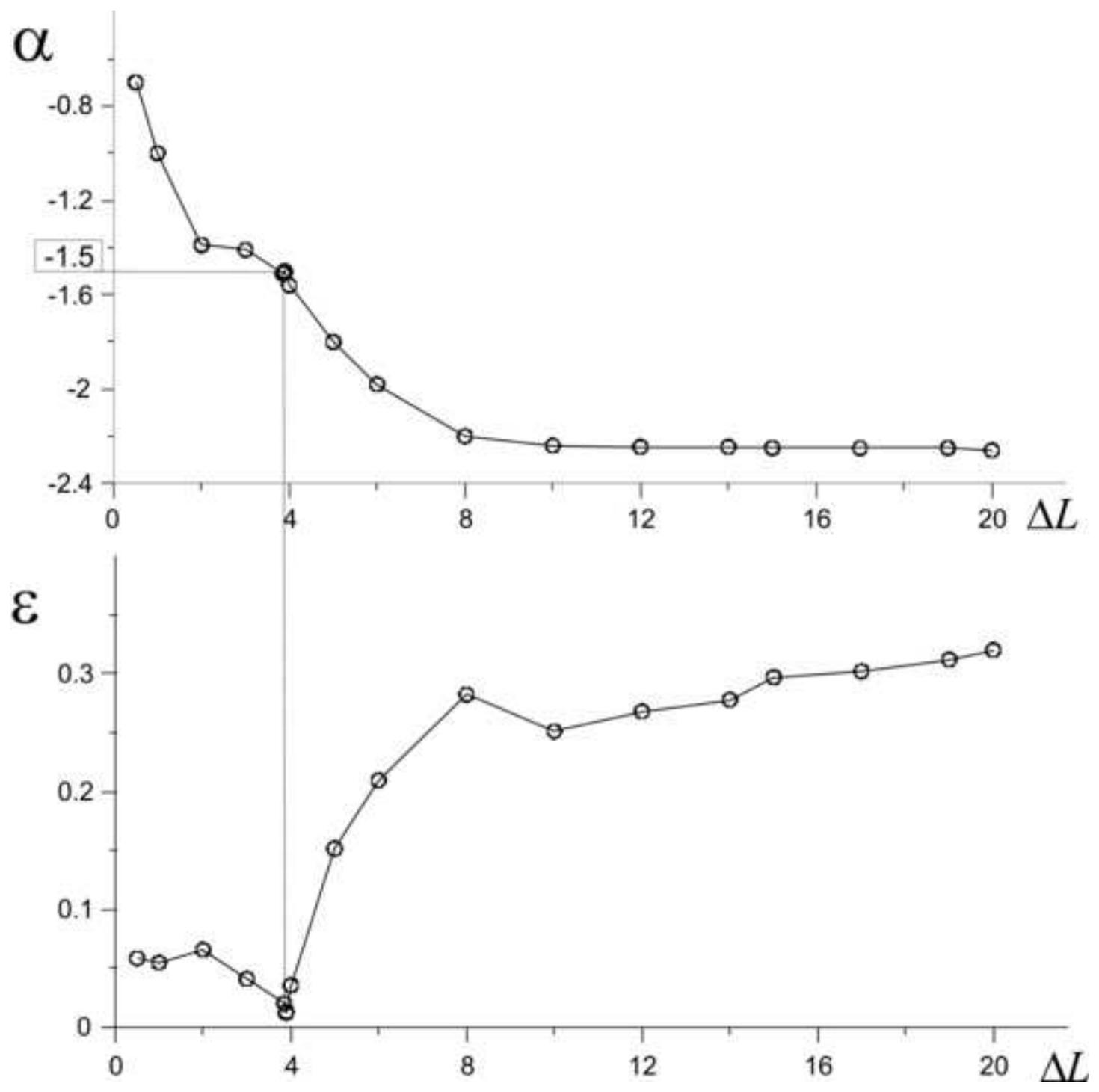

**Figure 3**

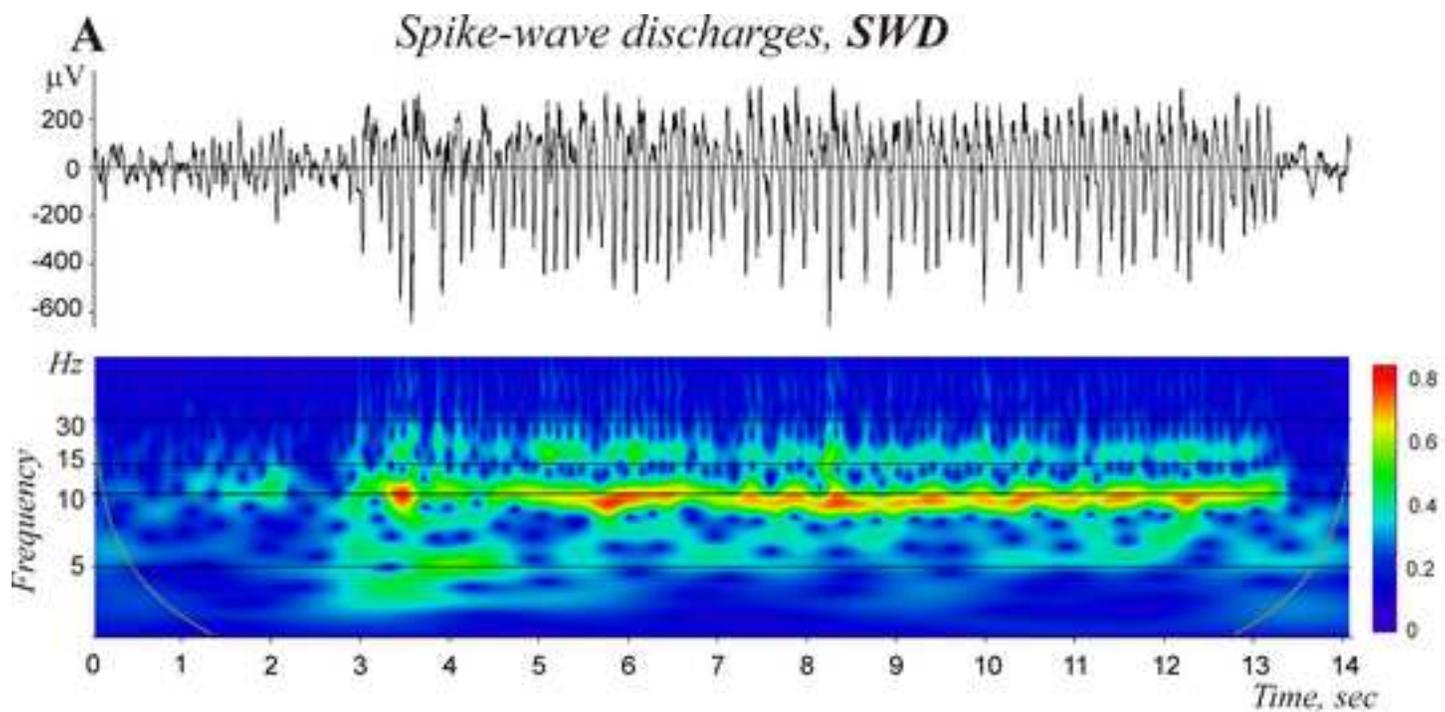

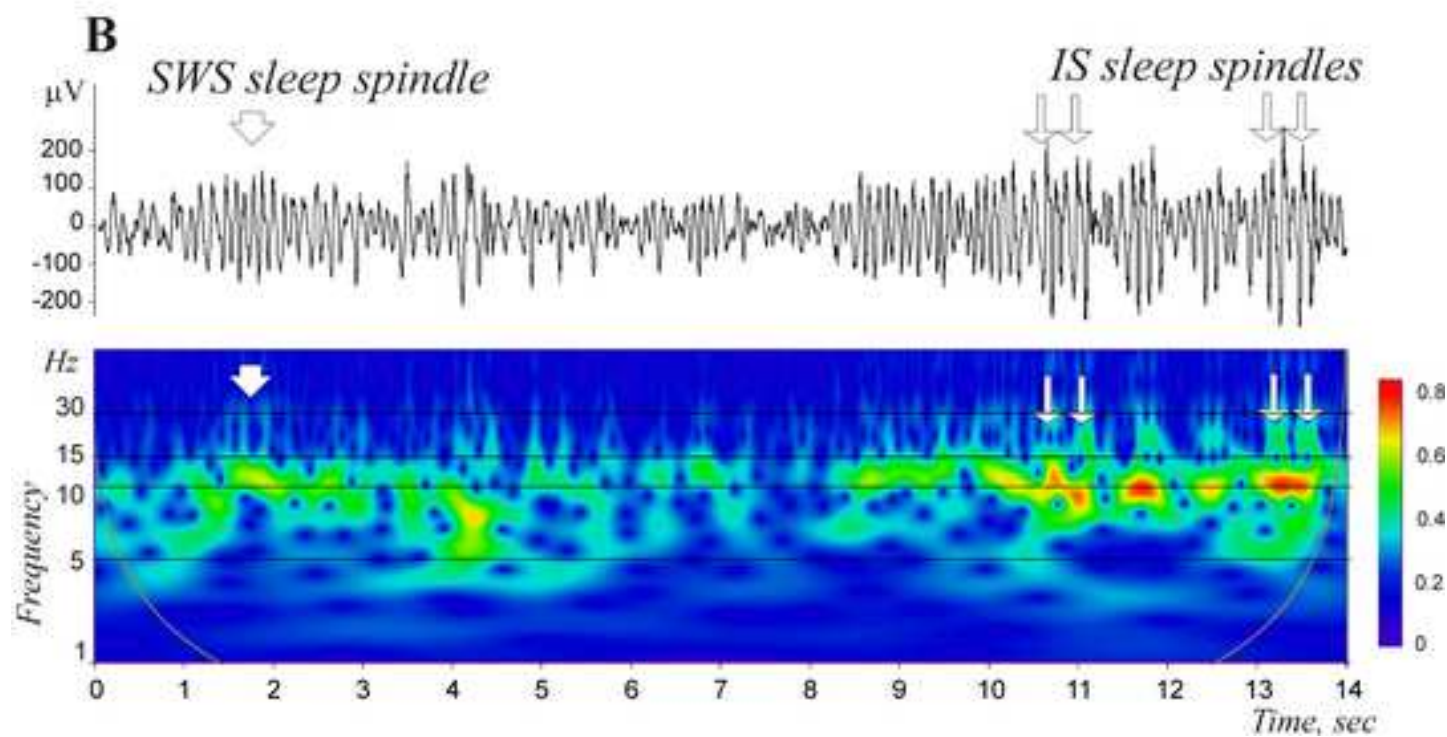

**Figure 4**

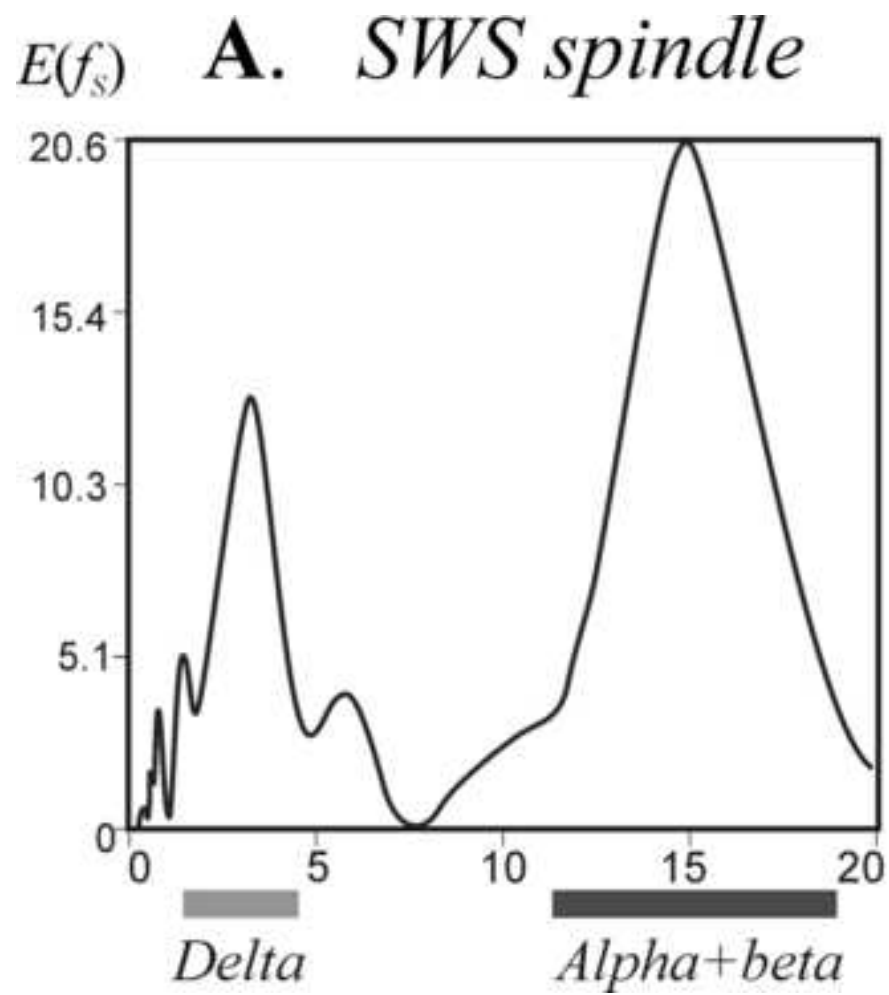 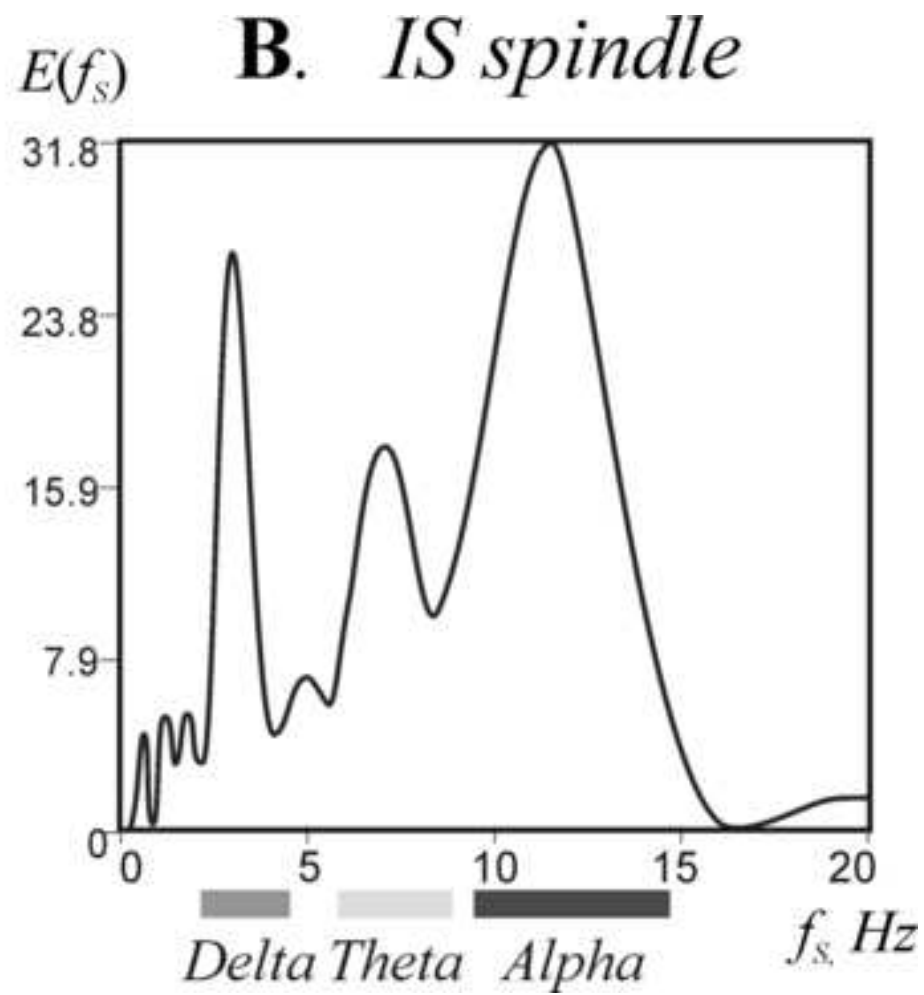

Figure 5

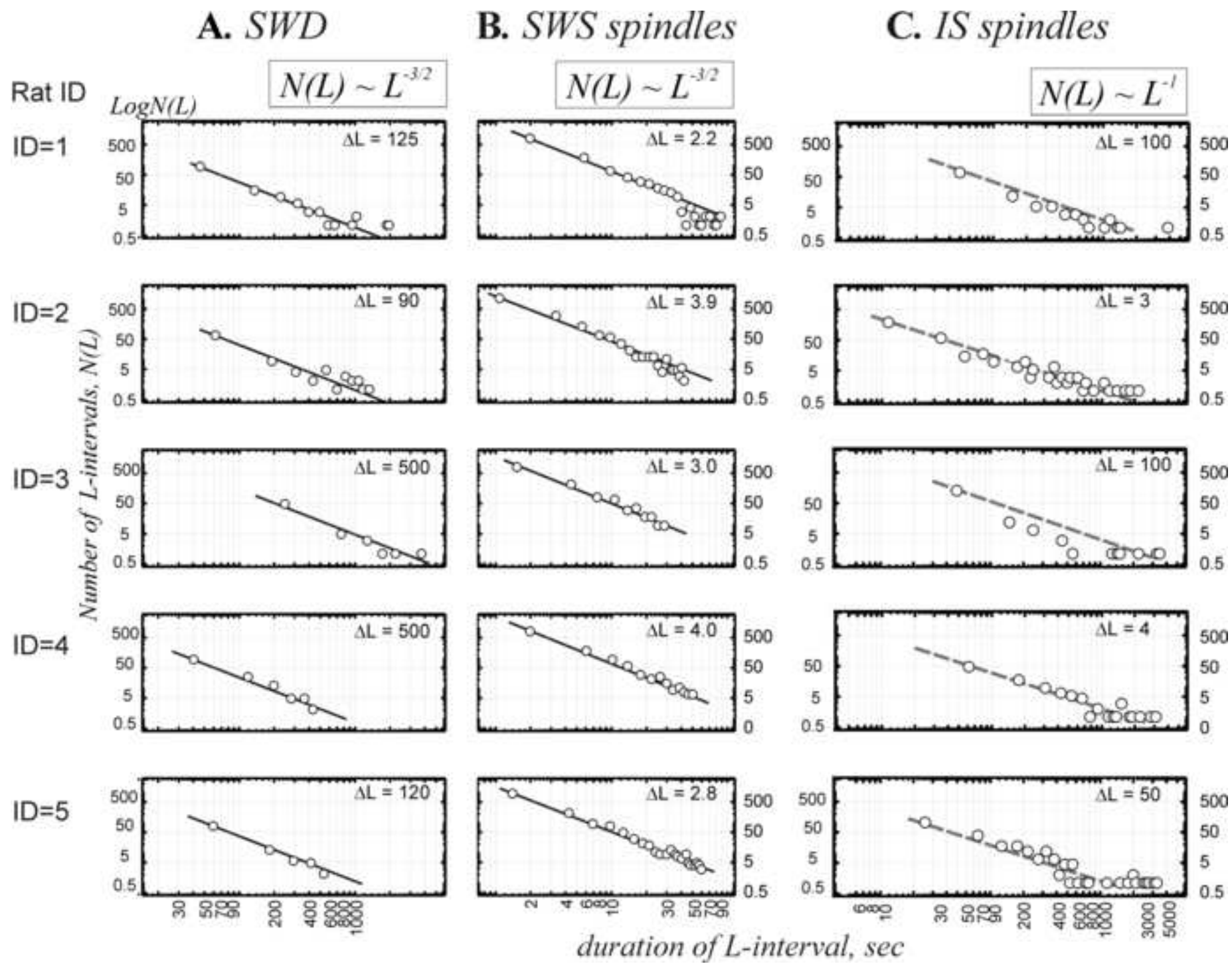